\documentclass[a4paper,11pt]{article}
\pdfoutput=1

\usepackage{jheppub}
\usepackage[T1]{fontenc}
\graphicspath{{img/}}

\usepackage[utf8x]{inputenc}
\usepackage[T1]{fontenc}
\usepackage{jheppub}
\usepackage{amsthm}
\usepackage{slashed}
\usepackage{mathtools}
\usepackage{float}
\usepackage{empheq}
\usepackage{bm}
\usepackage{framed}
\usepackage{comment}
\usepackage{xcolor}

\definecolor{Mblue}{rgb}{0.37,0.51,0.71}
\definecolor{Morange}{rgb}{0.88,0.61,0.14}
\definecolor{Mgreen}{rgb}{0.56,0.69,0.19}

\def\ss{\subsection}

\def\ie{\emph{i.e.} }

\def\R{\mathbb{R}}
\def\Z{\mathbb{Z}}

\def\cE{\mathcal{E}}

\def\cH{\mathcal{H}}
\def\cI{\mathcal{I}}

\def\cN{\mathcal{N}}

\def\fh {\mathfrak{h}}

\def\p{\partial}

\def\/{\over}

\def\rn{\rangle}
\def\ln{\langle}

\def\e{\epsilon}
\def\ve{\varepsilon}

\def\a{\alpha}

\def\d{\delta}

\def\g {\gamma}

\def\z{\zeta}

\def\l{\ell}

\def\n {\nabla}
\def\L{\Lambda}
\def\D{\Delta}
\def\G {\Gamma}

\def\r{\mathrm}

\def\_{\hspace{2cm}}
\def\-{\\\notag}
\def\={&=&}

\newcommand\be{\begin{equation}}
\newcommand\ee{\end{equation}}

\newcommand{\bea}{\begin{eqnarray}}
\newcommand{\eea}{\end{eqnarray}}

\newcommand{\bpm}{\begin{pmatrix}}
\newcommand{\epm}{\end{pmatrix}}

\newcommand{\bit}{\begin{itemize}}
\newcommand{\eit}{\end{itemize}}

\newcommand{\ben}{\begin{enumerate}}
\newcommand{\een}{\end{enumerate}}

\newcommand\bsp{\begin{split}}
\newcommand\esp{\end{split}}

\def\le{\left}
\def\ri{\right}

\def\l{\ell}

\def\qq{\qquad}

\title{Inflation and topology from the no-boundary state}

\author{Victor Godet}
\affiliation{LPTHE, Sorbonne Universit\'e, CNRS, 4 place Jussieu, 75005 Paris, France}
\emailAdd{vgodet@lpthe.jussieu.fr}

\abstract{
The no-boundary wavefunction for slow-roll inflation on the 3-sphere exponentially favors a small universe, in sharp disagreement with observations. We show that this problem is resolved by changing the spatial topology to the 3-torus. The sum over the ${\rm SL}(3,\mathbb{Z})$ family of geometries, computed using the theory of automorphic forms for ${\rm GL}(3)$, produces a wavefunction favoring a large inflating universe with ${\cal N} \gtrsim 250$ $e$-folds. We also compute corrections to the CMB power spectrum induced by  torus moduli fluctuations.}

\begin{document}
\maketitle
\flushbottom

\section{Introduction}\label{sec:intro}

The Hartle-Hawking proposal defines the wavefunction of the universe via a path integral over regular geometries with no past boundary \cite{Hartle:1983ai}. This proposal correctly predicts the Gaussian fluctuations observed in the CMB \cite{Halliwell:1984eu}. In the context of slow-roll inflation on the $S^3$ topology, it gives a probability
\be\label{S3prob}
|\Psi[S^3]|^2 \propto \r{exp}\Big({24\pi^2 M_p^4\/V(\phi_\ast)}\Big)
\ee
for the universe to start inflating at a potential value $V(\phi_\ast)$ in Planck units $M_p^{-2} = 8\pi G$. This exponentially favors a small universe, \ie few $e$-folds and large spatial curvature  \cite{Vilenkin:1987kf,Hartle:2007gi,Hartle:2008ng,Janssen:2020pii,Lehners:2023yrj,Maldacena:2024uhs}.

From a theoretical perspective, the $S^3$ topology is favored at the level of the classical action in the sum over topologies. However, it has been emphasized recently that using the inner product consistent with the Wheeler-DeWitt equation near $\cI^+$ \cite{Chakraborty:2023yed,Chakraborty:2023los}, the norm of the Hartle-Hawking state on $S^3$ vanishes \cite{Cotler:2025gui,Cotler:2026wlk}. This suggests that $\Psi[S^3]$ is actually a null state which should be removed from the physical Hilbert space. From this perspective, the Hartle-Hawking prescription actually predicts non-trivial spatial topology. The $S^2\times S^1$ wavefunction  \cite{Laflamme:1986bc,Conti:2014uda,Turiaci:2025xwi} gives $|\Psi[S^2\times S^1]|^2 \propto \r{exp}(16\pi^2 M_p^4/V(\phi_\ast))$. Although this topology dominates over the torus,  it suffers from the same problem as $S^3$ with slow-roll inflation.

The focus of this paper will be on the $T^3$ topology.  There, the wavefunction is actually an infinite sum over an $\r{SL}(3,\Z)$ family of geometries, all the possible ways of smoothly filling the 3-torus. After computing the sum and integrating over moduli,  the Hartle-Hawking norm takes the form
\be
(\Psi[T^3],\Psi[T^3]) \propto {M_p^4\/V(\phi_\ast) }~.
\ee
This gives a normalizable probability distribution on the number $\cN$ of $e$-folds given by $ P(\cN) = 2\e_V\, e^{-2\e_V \cN}$ where $\e_V$ is the slow-roll parameter \eqref{epsilonV}. This predicts 
\be
\cN \approx (2\e_V)^{-1} \gtrsim 250
\ee
using  $\e_V\lesssim 2\cdot 10^{-3}$  from the bound on the tensor-to-scalar ratio $r=16\e_V$ \cite{BICEP:2021xfz,Planck:2018jri}. This leads to a large universe consistent with observations. We will also see that the subleading terms in $\Psi[T^3]$ give corrections to the CMB power spectrum due to torus moduli fluctuations.

Quantum cosmology with toroidal topology has been considered \cite{Gowdy:1973mu,Moncrief:1980pq,Zeldovich:1984vk,Higuchi:1991tp,Hervik:2000ed,Linde:2004nz,Castro:2012gc,Guth:2025dal,Fornal:2026qfq}  but has received relatively little attention compared to $S^3$. Recently it has been shown that quantum cosmology on $T^d$ can be viewed as \emph{automorphic dynamics} \cite{Godet:2024ich,Godet:2025bju}, a physical incarnation of  the Langlands program for $\r{GL}(d)$. The powerful tools coming from number theory, such as spectral decompositions and trace formulas, make toroidal quantum cosmology highly computable. This paper demonstrates how these tools can be used to obtain definite predictions for cosmological observables.

\section{Toroidal slow-roll inflation}\label{sec:noboundary}

We consider gravity coupled to a scalar field
\be
S = \int d^4x \sqrt{-\hat{g}}\le( {1\/16\pi G} R- {1\/2}(\n\phi)^2 - V(\phi)\ri) - {1\/8\pi G}\int d^3u\sqrt{g}\,K~.
\ee
We focus on the slowly rolling regime where we are in a region of the potential with a small slow-roll parameter
\be\label{epsilonV}
\e_V= { M_p^2\/2}\Big({ V'\/ V}\Big)^2~.
\ee
We will think of inflation as happening between $\phi=\phi_\ast$ and $\phi=\phi_r$, after which reheating starts and the universe stops inflating.

\ss{Mini-superspace with inflaton}\label{sec:minisuperspace}

In the mini-superspace approximation, the metric takes the form
\be
ds^2 = \l_\r{dS}^2(-N(t)^2 dt^2 + g_{ij}(t)du^idu^j), \qq u_i\sim u_i+2\pi~,
\ee
and we  parametrize the spatial metric as $g_{ij} = T^{2/3} (g_z)_{ij}$ where
\be
T = \sqrt{\r{det}\,g},\qq g_z = {z\cdot z^t\/\r{det}(z)^{2/3}}
,\qq \r{det}\,g_z= 1~.
\ee
Here   the moduli of the torus are parametrized by 
\be\arraycolsep=3pt\def\arraystretch{1.1}
z=  \bpm 1& x_{2} & x_{3} \\ 0 & 1 &  x_{1} \\ 0 & 0 & 1 \epm\cdot   \bpm y_1 y_2 & 0 & 0\\ 0 & y_1 & 0 \\ 0 &0 & 1\epm =\bpm y_1 y_2 & y_1 x_2 & x_{3} \\ 0 & y_1 & x_{1} \\ 0 & 0 & 1 \epm \in \fh^3
\ee
which is an element of the generalized upper half-plane $\fh^3$ satisfying $\r{det}(z) = y_1^2 y_2$. We can write the metric on the torus as
\be
ds^2 = L_1^2 du_1^2+ L_2^2 (du_2+ x_2 du_1)^2 + L_3^2 (du_3+ x_1 du_2 + x_3 du_1)^2
\ee
where the lengths of the circles are given as
\be
L_1^3 = T y_1 y_2^2,\qq 
L_2^3 = {T y_1\/ y_2},\qq L_3^3 = {T\/y_1^2 y_2}
\ee
and the volume is
\be
T = L_1 L_2 L_3 = {1\/(2\pi)^3} \int d^3 u \sqrt{g}~.
\ee
We can compute the Einstein-Hilbert action of this spacetime. The Gibbons-Hawking-York term gives a contribution that cancels terms of the form $\p_t N$. After integrating out $N(t)$, we obtain the action of  a particle 
\be
S_\text{grav}=-\int dt\,\sqrt{-G_{ab}\dot{X}^a\dot{X}^b}~,
\ee
on an  auxiliary spacetime with metric
\be\label{auxmetric}
ds^2_\r{Aux} =G_{ab}dX^a dX^b={ (2\pi)^5 \l_\r{dS}^6V(\phi)\/3G}\le(  -dT^2+ {3\/4}T^2 ds^2(\fh^3)+12\pi G \,d\phi^2\ri)~.
\ee
We can always take the particle to have unit mass by absorbing it in the Weyl factor of the auxiliary metric. The metric on $\fh^3$ is the $\r{SL}(3,\R)$ invariant metric taking the form
\be\label{h3metric}
ds^2(\fh^3) = {dx_1^2\/y_1^2} + {dx_2^2\/y_2^2}+ {(dx_3-x_2 dx_1)^2\/y_1^2 y_2^2}+ {4\/3}\Big({dy_1^2\/y_1^2}+{dy_1 dy_2\/y_1 y_2}+{dy_2^2\/y_2^2}\Big)~.
\ee
The Wheeler-DeWitt equation then reduces to the   Klein-Gordon equation on \eqref{auxmetric}:
\be\label{KGauxWDW}
(G^{ab}\n_a\n_b -1)\psi(T,z)=0
\ee 
The Hilbert space then takes the form
\be
\cH = L^2(\r{SL}(3,\Z)\backslash \fh^3) \otimes\dots
\ee
where what appears is the Hilbert space of automorphic forms for $\r{GL}(3)$. There are two Casimir elements in $\r{SL}(3,\R)$ corresponding to two invariant differential operators:  $\D_1$ is the Laplacian on the metric \eqref{h3metric} and $\D_2$ is a cubic Casimir. The Langlands spectral decomposition gives a basis of $\cH$ in which $\D_1$ and $\D_2$ are diagonal  \cite{langlandsFunctionalEquationsSatisfied1976,Goldfeld_2006}
\begin{eqnarray}\label{LanglandsDec3}
\psi(z) &=& \text{const}+{1\/(4\pi i)^2}\int_{({1\/3})}ds_1\int_{({1\/3})}ds_2 \,(\psi, E_{s_1,s_2}) E_{s_1,s_2}(z) \-
&& \hspace{2cm}+ {1\/2\pi i }\sum_{j\geq 0} \int_{({1\/2})} ds \,( \psi, E_s^{(j)}) E_s^{(j)}(z) +\sum_{k\geq 1} \,(\psi,v_k)v_k(z)~.
\end{eqnarray}
which we write in terms of the Petersson inner product \eqref{Petersson} on $\fh^3$. The continuous spectrum consists of the Eisenstein series:  the minimal parabolic Eisenstein series $E_{s_1,s_2}$ and the maximal parabolic Eisenstein series $E_s^{(j)}$ twisted by the $\r{SL}(2,\Z)$ Maass forms $u_j$. The discrete spectrum is the set $\{v_k\}$ of $\r{SL}(3,\Z)$ Maass forms. These give the elementary wavefunctions for quantum cosmology, since they diagonalize the Wheeler-DeWitt equation.

\ss{Slow-roll solutions}\label{sec:geodesics}

Solutions of the Einstein equation are geodesics on the auxiliary spacetime. They project to geodesics on $\fh^3$. Due to the $\r{SL}(3,\R)$ symmetry, it is enough to study the geodesic where all coordinates are fixed except $y_1(t)$. These ``vertical'' geodesics will also be the ones relevant in the no-boundary solution.

The problem then reduces to studying a massive particle on a 3-dimensional auxiliary spacetime
\be\label{AuxReduced}
ds^2_\r{Aux}={ (2\pi)^5 \l_\r{dS}^6V(\phi)\/3G}\le(  -dT^2+ T^2  {dy_1^2\/y_1^2}+12\pi G \,d\phi^2\ri)~.
\ee
We will  take an exponential potential
\be\label{ExpPotential}
V(\phi) =V_0\, e^{-\sqrt{16\pi G\e_V} \,\phi}
\ee
expressed in terms of the slow-roll parameter $\e_V$. After some point $\phi=\phi_r$, the potential changes shape and the slow-roll approximation breaks down. We will only be interested in the slow-roll regime and give an exact solution for the potential \eqref{ExpPotential}.   Note that the exponential potential is ruled out by experiments. We use it as a solvable potential for the slow-roll regime, which will also give the leading order in $\e_V$ for a general potential.

The Einstein and scalar equations are then equivalent to the geodesic equation on the spacetime \eqref{AuxReduced}, which can be solved exactly. We are looking for solutions with the boundary conditions
\be
\lim_{t\to0}\,(T(t),y_1(t),\phi(t)) = (0,\infty,\phi_\ast) ,\qq \lim_{t\to+\infty} (T(t),y_1(t),\phi(t)) = (\infty, y_1, \infty)
\ee
This corresponds to a torus that starts at zero volume $T(0)=0$ and inflaton $\phi(0)=\phi_\ast$ and expands to infinite volume as the inflaton rolls. The asymptotic value of $y_1(t)$ is taken to be $y_1$.  The exact solution is given by
\be
\begin{dcases}
T(t) & =  c_1 t^{1\/2}(t+\a)^{3+\e_V\/2(3-\e_V)}~, \\
y_1(t) & = y_1 \sqrt{1+{\a\/t}}~, \\ 
\phi(t) & = \phi_\ast + {1\/2(3-\e_V)}\sqrt{\e_V\/\pi G}\, \log\Big(1+{t\/\a}\Big)~.
\end{dcases}
\ee
The real Lorentzian solutions should have $c_1,\a>0$. They always have a singularity where the volume $T$ vanishes at $t=0$, as was emphasized recently \cite{Guth:2025dal}. This is a difference from the Lorentzian solution on $S^3$ which is the regular dS$_4$ geometry.  In the next section, we will see that after slightly complexifying the solution, it does become regular with the geometry capping off smoothly at $T=0$, while the solution remains real at $T\to+\infty$. The resulting saddles thus  contribute to the no-boundary wavefunction.

\section{The Hartle-Hawking wavefunction}

\ss{No-boundary solutions}

We can study the behavior of the solution as $t\to0$. In this limit $T(t)\to0$ so the volume goes to zero and the torus becomes singular. If we expand the metric near $t=0$, we see that if we choose the value 
\be
c_1= {i\/\a^{3\/3-\e_V} (2\pi G \l_\r{dS}^2 V(\phi_\ast) (3-\e_V))^{3/2}}
\ee
the metric in the coordinate $\rho=\sqrt{t}$ takes the form
\be
ds^2= -{1\/2\pi G  \a V(\phi_\ast) (3-\e_V) }(d\rho^2 + \rho^2 du_3^2+\dots),\qq \rho\to 0
\ee 
which caps off smoothly at $\rho=0$. Moreover near $\rho\to0$, we have
\be
\phi = \phi_\ast+ \sqrt{\e_V\/\pi G} {\rho^2\/2\a(3-\e_V)} + O(\rho^4)
\ee
so that the regularity condition for the inflaton $\p_\rho\phi|_{\rho=0} =0 $  is also satisfied.

We can then compute the action and we find the contribution to the wavefunction
\be\label{T3saddlecontrib}
\Psi_0 =e^{-C_\ast\r{det}(z)}, \qq C_\ast  = {\pi(3+\e_V)\/3 G^2 (3-\e_V)^2 V(\phi_\ast)}~,
\ee
where we removed the universal divergent phase in the large $T$ limit. For $\e_V=0$, we recover the pure gravity answer
\be
C_\ast|_{\e\to0}  = {\pi\/9 G^2 V(\phi_\ast)} = {8\pi^2\l_\r{dS}^2\/27 G}
\ee
using that $V(\phi_\ast) = {\L\/8\pi G}$ in the constant potential case. 

\begin{figure}
	\centering
	\includegraphics[width=10cm]{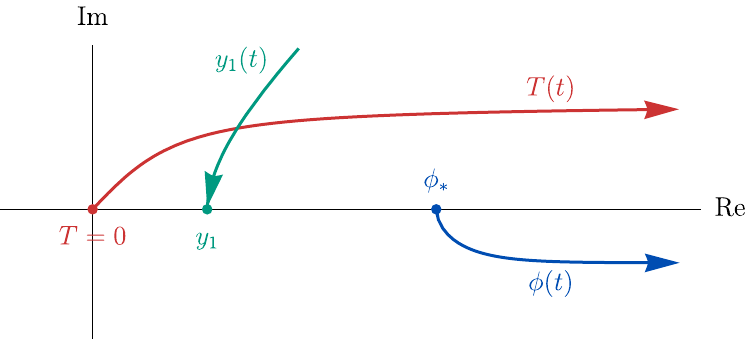}
	\caption{Profiles of the no-boundary solution for $(T(t),y_1(t),\phi(t))$ in the complex plane. The solution is complex and caps off smoothly at $T=0$ while becoming asymptotically real  as $T\to+\infty$. For this plot, we used $\e_V=0.1$.}\label{Fig:profiles}
\end{figure}

We see that $c_1$ is complex which makes $T(t)$ complex. To have a solution that is real asymptotically, we should choose $\a$ to be complex:
\be
 \a = i\ve \,e^{-i\pi\e_V/6}~,
\ee
where $\ve = |\a|>0$. The  phase of $\a$ is chosen to make the solution asymptotically real as can be seen by expanding the metric around $t=+\infty$. For illustration, the metric for $\e_V=0$ is explicitly
\be
ds^2 = {4 \l_\r{dS}^2\/9}\le( - {dt^2\/4t(t+i\ve)} + {(t+i\ve)^{2/3}\/\ve^{2/3}} \Big( y_1^2 y_2^2 du_1^2 + y_1^2 du_2^2 + {t\/t+i\ve} du_3^2\Big) \ri)~.
\ee 
This metric is complex in a way that makes it smooth at $t=0$ and becomes real in the $t\to+\infty$ limit. The complexification of the metric is similar to the $i\ve$ prescription in field theory. The metric is not isotropic at small $t$ but becomes isotropic at large $t$, which is the general expectation  \cite{Wald:1983ky}. For $\e_V=0.1$, the profile  for $(T(t),y_1(t),\phi(t))$ in the complex plane is plotted in Figure~\ref{Fig:profiles}.

Note that our solutions do not satisfy the proposed KSW criterion for allowable metrics in quantum gravity \cite{Kontsevich:2021dmb,Witten:2021nzp}, which has been used to constrain slow-roll inflation \cite{Hertog:2023vot,Lehners:2023pcn}. This criterion comes from imposing  positivity for the action of $p$-forms but given that the gravitational action is itself non-positive, it seems plausible that the necessary imaginary rotation would also affect the matter action, which would change  the criterion.

 Despite not satisfying KSW, the no-boundary geometries described in this paper lead  to a consistent wavefunction that gives normalizable probability distributions consistent with a large universe. Thus these geometries appear to be physically meaningful. The double-cone wormhole also violates KSW while having a clear physical interpretation as the ramp in the spectral form factor \cite{Chen:2023hra}. This suggests that the KSW criterion should be revisited or extended.

\ss{Sum over geometries}

The no-boundary  saddle discussed above gives the contribution
\be
\Psi_0(z) = \r{det}(z)^{1/2}\,e^{-C_\ast\,\r{det}(z)}
\ee
to the wavefunction, where we included the factor $ \r{det}(z)^{1/2}$ which comes from one-loop effects.\footnote{This is the right factor to include for the inner product to be the Petersson inner product \eqref{Petersson} due to the $bc$ ghost partition function \cite{Godet:2025bju}. Note that this factor has no real effect on our results.} Here  $C_\ast$ depends on the inflaton through $V(\phi_\ast)$. The Hartle-Hawking state on $T^3$ is the sum over all the ways to fill the $T^3$. This is a sum over $(n_1,n_2,n_3)\in \Z^3$ with the condition $\r{gcd}(n_1,n_2,n_3)=1$, corresponding to all the primitive combinations of the three circles. We can write this as
\be\label{HHT3inf}
\Psi_\r{HH}(z) =\tfrac12\sum_{\g\in P\backslash\r{SL}(3,\Z)} \Psi_0(\g z)
\ee
where $P$ is a maximal parabolic subgroup of $\r{SL}(3,\Z)$.   The sum over geometries can be viewed as a sum over vertical trajectories in the billiard $\r{SL}(3,\Z)\backslash \fh^3$. This perspective was emphasized in \cite{Godet:2024ich} in 3d gravity.

This Poincaré sum was already derived in \cite{Godet:2025bju}. Here we will review the main facts about it. The maximal parabolic Eisenstein series is defined as
\be
E_s(z)=\tfrac12\sum_{\g\in P\backslash\r{SL}(3,\Z)} \r{det}(\g z)^s~,
\ee
and we see that the Hartle-Hawking spectral representation is obtained directly from the Cahen-Mellin integral. We then obtain
\be
\Psi_\r{HH}(z) = -{4\pi \sqrt{\pi C_\ast}\/3\z(3)} + {1\/2\pi i}\int_{({1\/2})}ds\,\G(\tfrac12-s)\,C_\ast^{s-{1\/2}}\,E_s(z)~.
\ee
Here the constant term comes from a pole of the Eisenstein series that we pick up after shifting the contour to the spectral line $\r{Re}(s)={1\/2}$.

The pole structure is made manifest by rewriting the integrand in terms of $\cE_s(z) =\L(\tfrac32 s)E_s(z)$ where $\L(s) = \pi^{-s}\G(s)\z(2s)$ is the completed zeta function, since then $\cE_s(z)$ has only simple poles at $s=0,1$. This shows that the wavefunction has an expansion over the non-trivial zeros of the Riemann zeta function. It takes the form
\be\label{HHRiemann}
\Psi_\r{HH}(z) =- {4\pi \sqrt{\pi C_\ast}\/3\z(3)}  -\sum_\rho  {C_\ast^{ {\rho\/3}-{1\/2}}\G({1\/2}-{\rho\/3})\/3\,\pi^{-\rho/2}\G({\rho\/2})\z'(\rho)} \cE_{\rho\/3}(z)
\ee
where the Eisenstein series that appear here are special because one of the three terms in their constant Fourier expansion vanishes due to $s=\tfrac\rho3$ where $\z(\rho)=0$.

Finally, the $\z(3s)^{-1}$ factor in the integrand implies a different expression. From the Dirichlet expansion $\z(s)^{-1} = \sum_{m\geq 1}\mu(m)m^{-s}$, we obtain
\be
\Psi_\r{HH}(z) = \sum_{m\geq 1}\mu(m) m^{1/2} Z_\r{CFT}[ m^3 C_\ast ],\qq Z_\r{CFT} [C] ={3 C\/2\pi} \sum_{\vec{n}\in \Z^3} Q_z(\vec{n})^{3/4} e^{-C Q_z(\vec{n})^{3/2}}~.
\ee
This shows that the Hartle-Hawking wavefunction is a Möbius average of CFT partition functions on $T^3$. This partition function is defined as a sum over $\Z^3$ with the quadratic form \eqref{EpsteinzetaQuad}. One can think of   $Z_\r{CFT}$ as an ``integrable'' object,  a regularized version of the  partition function of the Liouville CFT$_3$  \cite{Levy:2018bdc,Kislev:2022emm} similar to a compact boson. In contrast, the Hartle-Hawking wavefunction is ``chaotic'' and we see that this chaos is implemented by a Möbius average, which has well-known pseudo-random properties \cite{sarnakThreeLecturesMobius,TaoBlogPost}. Together with the Riemann zeros expansion, this illustrates how  the sum over geometries for $T^3$ leads to number-theoretic pseudo-randomness in the no-boundary state. Further details on this are given in \cite{Godet:2025bju}.

\section{The size of the universe}
\label{sec:T3prob}

If the universe is a torus, its scale is probably much larger than the observable universe. Thus the torus moduli are not directly observable and should be integrated out, as in the no-boundary density matrix \cite{Ivo:2024ill}. Hence we should compute the Petersson norm
\be\label{Petersson}
(\Psi,\Psi) = \int_{\r{SL}(3,\Z)\backslash \fh^3} {dy_1 dy_2 dx_1 dx_2dx_3\/(y_1y_2)^3} |\Psi(z)|^2
\ee
in which the measure is fixed by $\r{SL}(3,\Z)$ invariance.

The Hartle-Hawking norm gives a probability distribution on the value $\phi_\ast$ at which inflation starts. From the spectral decomposition, we can compute the norm at leading order in $\e_V$ \cite{Godet:2025bju}
\be
P(\phi_\ast) =(\Psi_\r{HH},\Psi_\r{HH} ) ={2^8 \pi^6\/3^4 \z(3)} {M_p^4\/V(\phi_\ast)} + O(1)~.
\ee
It is simply  the absolute value squared of the constant term times $\r{vol}(\r{SL}(3,\Z)\backslash \fh^3) = {\z(3)\/4}$, while the Eisenstein contributions give an $\r{O}(1)$ term that we can ignore. This shows that after summing over geometries and integrating out shape moduli, the probability is just proportional to $1/V(\phi_\ast)$ with no exponential factor.   

We assume that reheating happens at $\phi=\phi_r$ beyond which we leave the slow-roll regime. So the universe inflates when $\phi$ rolls from $\phi_\ast$ to $\phi_r$.  From the exact solution, we can see that the spatial volume $T$ depends on $\phi$ so that at reheating it is given by
\be
T_r \approx c_1 \, e^{ 6\sqrt{\pi G\/\e_V} (\phi_r-\phi_\ast)}
\ee
This gives the direct relationship between $\phi_\ast$ and the number of $e$-folds:
\be
\cN = 2\sqrt{\pi G\/\e_V} \,(\phi_r-\phi_\ast)~.
\ee
Thus the Hartle-Hawking state leads to a normalizable probability distribution in $\cN$:
\be
P(\cN) =   e^{\sqrt{16\pi G\e_V}\, (\phi_\ast-\phi_r)}  = \r{exp}(-2\e_V\cN)~,
\ee
so that the normalized distribution is $dP = 2\e_V\, e^{-2\e_V\cN}\, d\cN$. This replaces the catastrophic exponent $2/A_s\sim 10^9$ for $S^3$ by the slow-roll-suppressed $2\e_V \lesssim 4\cdot 10^{-3}$ from the BICEP/Keck and Planck bound \cite{BICEP:2021xfz,Planck:2018jri}. The typical value of $\cN$ is then
\be
\ln \cN\rn = {1\/2\e_V} \gtrsim 250
\ee
which is comfortably above the observationally required $\cN \gtrsim 60$ \cite{Liddle:2003as,Planck:2018jri}. Moreover we can compute the probability 
\be
P(\cN > 60) = e^{-120\,\e_V} \gtrsim 79\%
\ee
so a sufficiently long inflationary phase is the generic outcome on $T^3$.

\section{Corrections to the CMB}

The Hartle-Hawking state defines a probability distribution $|\Psi_\r{HH}(z)|^2$ on the torus shape. The  leading piece $\propto 1/V(\phi_\ast)$  is moduli-independent but the  subleading corrections depend non-trivially on $z$. Cosmological observables sensitive to the spatial shape should therefore be averaged against $|\Psi_\r{HH}(z)|^2$. We illustrate this with a simple example, the angular CMB power spectrum.

Decomposing the temperature anisotropy in spherical harmonics according to ${\d T(\hat n)\/T} = \sum_{\l m} a_{\l m} Y_{\l m}(\hat n)$, the angular power spectrum is the rotationally averaged variance of the multipole coefficients,
\be
C_\l = {1\/2\l+1}\sum_{m=-\l}^{\l} \ln |a_{\l m}|^2\rn~.
\ee
In the Sachs-Wolfe approximation, $\d T/T$ on the celestial sphere is sourced by the primordial curvature perturbation $\z$ at the last-scattering surface. With a scalar power spectrum $P_\z(k)\propto k^{n_s-4}$, this gives on $\R^3$
\be
C_\l^{(0)} = {1\/2\pi} \int d^3 \vec{k}\, j_\l(\chi_\r{rec}\, k)^2\, k^{n_s-4}~,
\ee
where $k=|\vec{k}|$ and  $j_\l$ is the spherical Bessel function, and $\chi_\r{rec}$ is the comoving distance to recombination.\footnote{We choose the normalization here so that for $n_s=1$, we have $C_\ell^{(0)} =1/(\l(\l+1))$.} On $T^3$, the spatial momenta are discrete and the integral becomes a sum over the dual lattice \cite{PhysRevLett.71.20,Scannapieco:1998cz,COMPACT:2023rkp}
\be
C_\l(z) = {1\/2\pi} \sum_{\vec n\in \Z^3\setminus\{0\}} j_\l(X k_{\vec n})^2\, k_{\vec n}^{n_s-4}, \qq X = {\chi_\r{rec}\/T^{1/3}}~,
\ee
where $k_{\vec n}$ is the magnitude of the wavevector labelled by $\vec n$ on the unit-volume torus of shape $z$, and $X$ measures the horizon size in units of the torus scale. The flat-space integral is recovered as $X\to 0$.

The quantity $C_\l(z)$ is a function of the torus moduli $z$ and is invariant under $\r{SL}(3,\Z)$. Thus we should be able to decompose it in the spectral decomposition \eqref{LanglandsDec3}. The way to do this is similar to the computation of the compact boson partition function on $T^d$, see \cite{Godet:2025bju} for more details. First we define the Epstein zeta function 
\be\label{EpsteinzetaQuad}
2 \z_z(s) =  \sum_{\vec{n}\in  \Z^3\setminus\{0\}} Q_z(\vec{n})^{-s}= \sum_{\vec{n}} k_{\vec{n}}^{-2s}~,\qq Q_z(\vec{n}) =  \vec{n}\cdot g_z^{-1} \cdot \vec{n}~,\qq k_{\vec{n}} = \sqrt{Q_z(\vec{n})}~,
\ee
associated to the quadratic form constructed from the inverse torus metric $g_z^{-1}$. It is related to the completed maximal parabolic Eisenstein series via
\be
\cE_{1-s}(z)= \pi^{-\tfrac{3}{2}s}\G(\tfrac{3}{2}s) \z_z(\tfrac{3}{2}s)~.
\ee
To write $C_\l(z)$ in terms of the Epstein zeta function, we  first write the Mellin transform
\be
j_\l(x)^2 = {\pi\/2x} {1\/2\pi i}\int_{i\R}ds {\G(-s)\G(2s+2+2\l)\/\G(s+\l+{3\/2})^2\G(s+2+2\l)} (\tfrac12 x)^{1+2\l +2s}
\ee
which follows from $j_\l(x) = \sqrt{\pi\/2x} J_{\l+{1\/2}}(x)$.  For convergence we should first shift the contour so that the Epstein sum is convergent. Then after replacing the Epstein zeta function with the Eisenstein series, we shift back the contour to the spectral line. This picks up a pole of the completed Eisenstein series which produces a constant term
\be
C_\l(z) = C_\l^{(0)} + {1\/2\pi i}\int_{({1\/2})} ds \,\rho_\l(s) E_s(z)~.
\ee
\begin{figure}
	\centering
	\includegraphics[width=15cm]{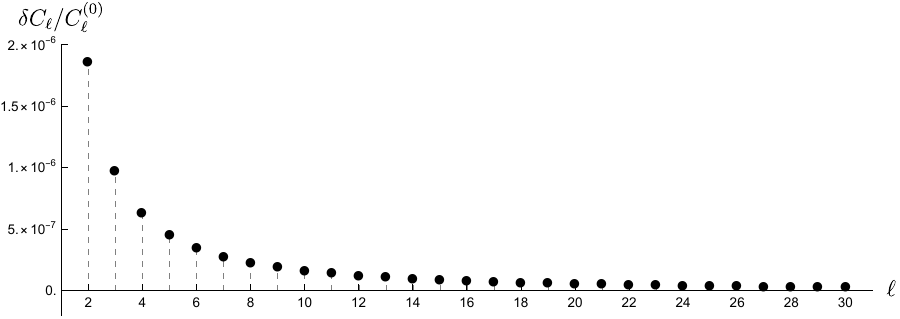}
	\caption{Correction to the CMB power spectrum due to torus moduli fluctuations in the Hartle-Hawking wavefunction. The plot is done for the values $C_\ast = 2\cdot 10^{11}, X=3, n_s=0.965$. We see that the effect is stronger at small $\ell$. However it is suppressed in $C_\ast^{-1/2}$ so that the correction is of order $10^{-6}$ of the leading value, which makes these quantum gravity fluctuations too small to be directly detectable.}\label{Fig:CMB}
\end{figure}
The constant term is just the flat $\R^3$ result which corresponds to the infinite volume limit:
\be
C_\l^{(0)} = {\sqrt{\pi} X^{1-n_s}\/2} {\G({3-n_s\/2})\G({2\l-1+n_s\/2})\/\G({4-n_s\/2})\G({5+2\l-n_s\/2})} ,\qq C_\l^{(0)}|_{n_s\to 1} = {1\/\l(\l+1)}~.
\ee
The second term contains the dependence on the torus shape, and it has only a contribution of the maximal parabolic Eisenstein series with density
\be
\rho_\l(s) = {3 \pi X^{1-n_s} \/4} (X/\pi)^{3s} {\G({3s\/2}) \G({3-n_s+3s\/2})\G({2\l+n_s-1-3s\/2})\/\G({3-3s\/2})\G({4-n_s+3s\/2})\G({2\l-n_s+5+3s\/2})} \z(3s)~.
\ee
We can then compute the expectation value in the HH state
\be
\ln C_\l\rn_\r{HH}  = {\int d^\ast z\, |\Psi_\r{HH}(z)|^2 C_\l(z) \/\int d^\ast z\,|\Psi_\r{HH}(z)|^2 }
\ee
where $ d^\ast z$ denotes the Petersson measure in \eqref{Petersson}. At large $C_\ast$, we can expand
\be
|\Psi_\r{HH}(z)|^2 = {16\pi^3  C_\ast\/9 \z(3)^2} - {4\sqrt{\pi}\/3\z(3)} \int_\R d\nu \,\G(-i\nu) C_\ast^{{1\/2}+i\nu} E_{{1\/2}+i\nu}(z) + O(1)
\ee
We then use the fact that $(E_{{1\/2}+i\mu},E_{{1\/2}+i\nu}) = 2\pi \d(\mu-\nu)$ to obtain the correction
\be
\d C_\l = \ln C_\l\rn_\r{HH}-C_\l^{(0)} = {1\/2\pi i }\int_{({3\/2})^-}ds\, K(s) \z(s)
\ee
which is expressed as a Mellin-Barnes integral of the Riemann zeta function on the line $\r{Re}(s)={3\/2}-\e$ with kernel
\be
K(s) =-  {24\/\sqrt{\pi}}  (2X)^{1-n_s}\big(\pi C_\ast^{1/3} /X\big)^{-s} {\r{cos}({\pi s\/2})\G({s\/3}-{1\/2})\G(s-1)\/\r{cos}({\pi (n_s-s)\/2})\G(s+4-n_s)} \prod_{m=1}^\l {2m-3+n_s-s\/2m+3-n_s+s} ~.
\ee
The correction is suppressed by $X^{(5-2n_s)/2} C_\ast^{-1/2}$. The experimental bounds from Planck \cite{Planck:2015gmu} for the cubic torus give $\pi L > 0.97 \chi_\r{rec}$. This implies that $X \lesssim 3.24$.
The value of $C_\ast$ follows from the scalar amplitude $A_s$ and the slow-roll parameter $\e_V$. Using $V(\phi_\ast) =24\pi^2 M_p^4 A_s\,\e_V$, we obtain $ C_\ast = \frac{8\pi}{27A_s\e_V} \gtrsim 2\times 10^{11}$ using $A_s\approx 2.1\times 10^{-9}$ \cite{Planck:2018vyg} and the  bound $\e_V\lesssim 2\times 10^{-3}$ mentioned above. Thus we find that the correction to the rotationally averaged spectrum is at most of order \(10^{-6}\) relative to the constant contribution, see Figure~\ref{Fig:CMB}. This is far below cosmic variance, so this particular observable is not expected to be detectable. The full anisotropic covariance would also be interesting to compute as it should contain finer predictions from toroidal topology. Our goal in this section was not to derive a detectable prediction, as the quantum gravity effects considered here are expected to be very small, but rather to show that the formalism is predictive.

\section{Discussion}

The no-boundary wavefunction for slow-roll inflation is inconsistent with a large universe with $S^3$ or $S^2\times S^1$ topology. The fact that the $T^3$ wavefunction leads to a normalizable probability distribution implying a large number of $e$-folds can be viewed as evidence for the $T^3$ topology.

The question of which topology dominates   depends on the choice of integration contour in the gravity path integral \cite{Gibbons:1978ac,Halliwell:1989dy,DiazDorronsoro:2017hti,Witten:2021nzp}, an active topic of current research. On the Lorentzian contour \cite{Feldbrugge:2017kzv,Feldbrugge:2017fcc,Feldbrugge:2017mbc,Loges:2022nuw,Marolf:2022ybi,Held:2026huj,Kolanowski:2026gii}, the integrand is a pure phase and steepest descent makes the contribution of every saddle-point exponentially suppressed: for $S^3$, the exponent in \eqref{S3prob} gets inverted. The $T^3$ saddles \eqref{T3saddlecontrib} are already suppressed, and the sum over the infinite mapping class group $\r{SL}(3,\Z)$ converts the exponential into the power law $|\Psi[T^3]|^2 \propto 1/V(\phi_\ast)$. Thus, on the Lorentzian contour, the $T^3$ topology dominates due to its large mapping class group, and the no-boundary wavefunction becomes a normalizable probability distribution predicting large inflating toroidal universes.

Changing the topology from $S^3$ to $T^3$ makes the mini-superspace dynamics much richer \cite{Godet:2024ich,Godet:2025bju}. The Hilbert space of quantum cosmology acquires a tensor factor  of  the space of automorphic forms $L^2(\r{SL}(3,\Z)\backslash \fh^3)$. The Langlands spectral decomposition for $\r{GL}(3)$ gives the basis of  physical wavefunctions and cosmological observables such as the Hartle-Hawking norm are computed by  trace formulas. This also leads to an infinite number of new physical symmetries: the $\r{SL}(3,\Z)$ Hecke operators which provide an integrable arithmetic structure.  The wavefunction $\Psi_\r{HH}$ is naturally viewed as a large constant term with small fluctuations controlled by the non-trivial zeros of the Riemann zeta function.

Experimentally, the $T^3$ topology is consistent with the observed flatness of the universe. Searches for cosmic topology in the CMB \cite{Cornish:1997ab,Cornish:2003db,Planck:2013okc,Planck:2015gmu,Aurich:2021ofm,PhysRevLett.71.20} have so far found no evidence; this effort has recently been revived by the COMPACT collaboration \cite{COMPACT:2022gbl,COMPACT:2023rkp,COMPACT:2024dqe,COMPACT:2026vdj}, which proposes new methods to detect non-trivial topology even when the topology scale is comparable to or larger than the observable horizon. Moreover, several persistent CMB anomalies at low multipoles \cite{Schwarz:2015cma,deOliveira-Costa:2003utu,Herold:2025mro} and recent discussions of statistical-isotropy violation \cite{Hajian:2003ic,Jones:2023ncn,Guth:2026qiu} would find a natural explanation if the universe has non-trivial topology. The  $T^3$ topology also leads to a moduli-dependent Casimir energy with potential phenomenological consequences \cite{COMPACT:2026vdj,Fornal:2026qfq}.

If the universe is a torus, its scale is likely much larger than our horizon, which would make direct detection difficult. We have shown that torus moduli fluctuations in the no-boundary state lead to corrections in cosmological observables such as the CMB power spectrum, but these quantum gravity effects are very small and unlikely to be detectable.  More promisingly, toroidal topology changes the inflationary dynamics and should affect many observable quantities, such as the number of $e$-folds demonstrated in this paper. The connection to number theory makes toroidal quantum cosmology unusually tractable, and we expect this to  yield many further predictions.

\acknowledgments We acknowledge fruitful discussions during the workshop ``Physics and automorphic $L$-functions: gravity, conformal field theory and number theory", organized at the Isaac Newton Institute for Mathematical Sciences (Cambridge) from 8--10 April 2026, and during the workshop ``Observers, wormholes and complex saddles in cosmology", organized at the Bernoulli Center for Fundamental Studies (EPFL, Lausanne) from 18--22 May 2026.

\bibliographystyle{JHEP}
\bibliography{refs}

\end{document}